\documentclass[12pt]{article}

\usepackage[top=1in, bottom=1in, left=1in, right=1in]{geometry}
\usepackage{natbib}
 \bibpunct{(}{)}{;}{a}{,}{,}
\usepackage{bm}
\usepackage{amsmath, amsthm}
\usepackage{bbm}
\usepackage{graphicx}
\usepackage{pbox}
\usepackage{booktabs}
\usepackage[colorlinks=true,citecolor=blue,linkcolor=blue]{hyperref}
\newtheorem{theorem}{Theorem}
\newtheorem{corollary}{Corollary}
\newtheorem{lemma}{Lemma}
\usepackage{rotating}
\usepackage{bm}
\usepackage{caption}
\usepackage{setspace}
\newcommand{\calC}{\mathcal{C}}

\def\E{E}

\newcommand{\N}{\mathcal{N}}

\newcommand{\Y}{Y}
\newcommand{\W}{W}
\newcommand{\Z}{Z}
\newcommand\indep{\protect\mathpalette{\protect\independenT}{\perp}}
\def\independenT#1#2{\mathrel{\rlap{$#1#2$}\mkern2mu{#1#2}}}
\newcommand{\proba}{\mathbbm{P}}
\newcommand{\var}{\operatorname{var}}
\newcommand{\defeq}{=}

\begin{document}

\title{Testing for arbitrary interference on experimentation platforms}

\author{Jean Pouget-Abadie$^\dagger$, Guillaume Saint-Jacques$^\mathsection$,
Martin Saveski$^\ddagger$,  \\ Weitao Duan$^\mathsection$, Ya Xu$^\mathsection$,
Souvik Ghosh$^\mathsection$,  Edoardo M. Airoldi$^\ast$\and $^\dagger$\,Google
Research, ~ $^\ddagger$\,MIT, ~ $^\mathsection$\,LinkedIn, \and  $^\ast$ Fox School of Business, Temple
University}

\maketitle

\begin{abstract}
Experimentation platforms are essential to modern large technology companies, as
  they are used to carry out many randomized experiments daily.  The classic
  assumption of no interference among users, under which the outcome of one user
  does not depend on the treatment assigned to other users, is rarely tenable on
  such platforms. Here, we introduce an experimental design strategy for testing
  whether this assumption holds. Our approach is in the spirit of the
  Durbin--Wu--Hausman test for endogeneity in econometrics, where multiple
  estimators return the same estimate if and only if the null hypothesis holds.
  The design that we introduce makes no assumptions on the interference model
  between units, nor on the network among the units, and has a sharp bound on
  the variance and an implied analytical bound on the type I error rate.   We
  discuss how to apply the proposed design strategy to large experimentation
  platforms, and we illustrate it in the context of an experiment on the
  LinkedIn platform.
\end{abstract}

\newpage
\tableofcontents
\newpage

\section{Introduction}

Applications of the potential outcomes approach to causal inference
\citep{rubin1974estimating,imbens2015causal} often rely on the assumption of no
interference among the units of analysis: the outcome of
each unit does not depend on the treatment assigned to other units. This fact is
formalized by the stable unit treatment value assumption
\citep[p. 475]{rubin1990formal}.  However, in applications where social
information among the units of analysis is recorded, this is often untenable,
and  classical results in causal inference may no longer hold true
\citep{karwa2017aa}.  Examples of causal analyses where interference is present
are numerous include education policy~\citep{hong2012evaluating}, viral
marketing campaigns~\citep{aral2011creating, eckles2016estimating}, and social
networks and healthcare \citep{shakya2016aa}. In these examples, for instance,
the difference-in-means estimator computed under Bernoulli randomization is no
longer guaranteed to be an unbiased estimator of the average treatment effect.

Significant efforts to extend the theory of causal inference to scenarios where
the stable unit treatment value assumption does not hold have been made. A
popular approach to minimizing the effects of interference, cluster-based
randomized designs, have been extensively studied, spanning from the early work
\citep{cornfield1978symposium, commit1991community, donner2004pitfalls} to more
recent contributions \citep{ugander2013graph, eckles2014design}. Multi-level
designs where treatment is applied with different proportions across the
population, known as randomized saturation designs \citep{hudgens2008toward,
tchetgen2012causal}, are also an important tenet of the literature on improving
causal estimates under interference, having been applied to vaccination trials
\citep{datta1999efficiency} and more recently to voter-mobilization campaigns
\citep{sinclair2012detecting}. See \citet{baird2014designing} for more
references. More recent literature has developed around various assignment
strategies and estimators, beyond cluster-based randomized design or multi-level
designs with some guarantees under specific models of
interference~\citep{backstrom2011network, katzir2012framework,
toulis2013estimation, manski2013identification, choi2014estimation,
basse2015optimal, gui2015network}.

Although dealing with interference for the purpose of estimating causal effects
is often important in  causal inference problems where  interference is due to a
network, a more fundamental need is to be able to detect interference in a given
experiment.  \citet{rosenbaum2007interference} was the first to formulate two
sharp null hypotheses that imply the stable unit treatment value assumption does
not hold. Under these restricted null hypotheses, the exact distribution of
network parameters is known. More recent work \citep{aronow2012general,
athey2015exact} explicitly tackles testing for the non-sharp null that the
stable unit treatment value assumption holds, by considering the distribution of
network effect parameters for a subpopulation of the graph under this
assumption. In contrast to these post-experiment analysis methods, this paper
suggests an experimental design test for interference.

We introduce a multilevel experimental design strategy that
allows the experimenter to test whether interference is present. This
design-centric approach differs from prior work, which focuses on applying
post-experiment analysis methods to classical experimental designs, such as the
Fisher randomization test. Our approach is in the spirit of the
Durbin-Wu-Hausman test for endogeneity in econometrics
\citep{hausman1978specification}, where multiple estimators return the same
estimate if and only if the null hypothesis holds.  The design that we introduce
makes no assumptions on the interference model between units, nor on the network
among the units, and comes with a sharp bound on the variance (under some
conditions), and an implied analytical bound on the type I error rate.  Most
importantly, the proposed design is non-intrusive: it allows the experimenter to
analyse the experiment in a classical way, for example to conduct an analysis of
the treatment effect with the standard assumptions, at the cost of reduced
power.

In Section~\ref{sec:methodology}, we discuss the methodology for the
experimental design. In Section~\ref{sec:theory}, we provide guarantees on the
type I and II error of our suggested test.  In Section~\ref{sec:practical}, we
discuss modifications to the suggested experimental design to meet practical
constraints. We illustrate the effectiveness of our design on a small-scale
simulation study in Section~\ref{sec:simulation}. Finally, in
Section~\ref{sec:results}, we present some of the results obtained for an
experiment launched in August 2016 on LinkedIn's experimentation platform using
our suggested framework.

\section{Methodology}\label{sec:methodology}

\subsection{Complete randomization and cluster-based
randomization}\label{sec:candidate_assign}

Consider $N$ experimental units and a possible intervention on these units.
Each unit's potential outcome $Y_i$ is a function of the entire
assignment vector $\Z \in (0,1)^N$ of units to one of two possible cases. If
$Z_i = 1$, unit $i$ is treated and given the intervention. Otherwise, $Z_i = 0$
indicating that unit $i$ is placed in control. The causal estimand of interest
is the total treatment effect:
\begin{equation*}
\textstyle
  \tau \defeq \frac{1}{N} \sum_{i=1}^N \left\{ Y_i(\Z = 1) - Y_i(\Z = 0) \right\}.
\end{equation*}
For any vector $u \in \mathbbm{R}^N$, let $\bar{u} = \frac{1}{N} \sum_{i
=1}^N u_i$ and $\sigma^2(u) = \frac{1}{N -1} \sum_{i=1}^N (u_i - \bar{u})^2$.
The total treatment effect can be rewritten: $\tau \defeq \overline{\Y(1)} -
\overline{\Y(0)}$.
Two popular experimental designs, the completely randomized design and the
cluster-based randomized design, provide unbiased estimates of the total
treatment effect.  In a completely randomized experiment,the
assignment vector $\Z$ is sampled uniformly at random from the set $\{z \in (0, 1)^N:
\sum z_i = n_t\}$, where  $n_t$ is the number of units assigned to treatment and
$n_c \defeq N - n_t$ is the number of units assigned to control.  The
difference-in-means estimator is $ \hat \tau_{cr} \defeq \overline{\Y_t} -
\overline{\Y_c}$, where $\Y_t \defeq (Y_i : Z_i = 1)$ is the outcome vector of
all units in treatment and $\Y_c \defeq (Y_i : Z_i = 0)$ is the outcome vector
of all units in control. Let $S_t \defeq \sigma^2\{\Y(1)\}$, $S_c \defeq
\sigma^2\{\Y(0)\}$ be the variances of the two potential outcomes under the stable
unit treatment value assumption and let $S_{tc} \defeq \sigma^2\{\Y(1) -
\Y(0)\}$ be the variance of the differences of the potential outcomes. The
following result is widely known in the causal inference literature.
\begin{lemma}
When the stable unit treatment value assumption holds, the expectation and
  variance of the difference-in-means estimator $\hat \tau_{cr}$ under a
  completely randomized design are
  \begin{align*}
    \E_{\Z} \left(\hat \tau_{cr}\right) & = \tau \\
    \sigma_{cr}^2 \defeq \var_{\Z} \left( \hat \tau_{cr} \right) &  =
    \frac{S_t}{n_t} + \frac{S_c}{n_c} - \frac{S_{tc}}{N}.
  \end{align*}
\end{lemma}

In a cluster-based randomized assignment, the randomization is over clusters of
units, rather than individual units.  Supposing that each experimental unit
is assigned to one of $M$ clusters, the cluster assignment vector $z$ is sampled
uniformly at random from $\{v \in (0, 1)^M: \sum v_i = m_t\}$, assigning units
in cluster $\calC_j$ to the corresponding treatment: $Z_i = 1 \Leftrightarrow
z_j = 1$ if $i \in \calC_j$, where $m_t$ is the number of clusters assigned to
treatment and $m_c \defeq M - m_t$ is the number of clusters assigned to
control. Let $\Y^+$ be the vector of aggregated potential outcomes, defined
as $Y^+_j \defeq \sum_{i \in \calC_j} Y_i$, the sum of all outcomes within
cluster $\calC_j$.  The Horvitz--Thompson estimator is defined as $\hat
\tau_{cbr} \defeq M/N \left( \overline{\Y^+_t} - \overline{\Y^+_c}
\right)$, where  $\Y^+_t \defeq (\Y^+_j : z_j = 1)$ is the cluster-level
outcome vector of all treated clusters and  $\Y^+_c \defeq (\Y^+_j : z_j = 0)$
is the cluster-level outcome vector of all clusters in the control bucket.
Let $S^+_t \defeq \sigma^2\{\Y^+(1)\}$, $S^+_0 \defeq \sigma^2\{\Y^+(0)\}$ be
the variance of the two aggregated potential outcomes under the stable unit
treatment value assumption and $S^+_{tc} \defeq \sigma^2\{\Y^+(1) - \Y^+(0)\}$
be the variance of the difference of the aggregated potential outcomes. The
following result is widely known in the causal inference literature.
\begin{lemma}
  \label{lem:cbr}
When the stable unit treatment value assumption holds, the expectation and
  variance of the Horvitz--Thompson estimator $\hat \tau_{cbr}$ under a
  cluster-based randomized design are
  \begin{align*}
    \E_{\Z} \left(\hat \tau_{cbr}\right) & = \tau, \\
    \sigma_{cbr}^2 \defeq
    \var_{\Z} \left( \hat \tau_{cbr} \right) & = \frac{M^2}{N^2} \left(
    \frac{S^+_t}{m_t} + \frac{S^+_c}{m_c} - \frac{S^+_{tc}}{M} \right).
  \end{align*}
\end{lemma}
Lemma~\ref{lem:cbr} does not require the clusters to be of equal size. However,
this assumption will be required for the hierarchical design presented in
Section~\ref{sec:hier}.  When the stable unit treatment value assumption holds,
$\hat \tau_{cr}$ and $\hat \tau_{cbr}$ are unbiased estimators of the total
treatment effect under their respective randomized designs.  However, when the
stable unit treatment value assumption does not hold, these result are no longer
guaranteed and the estimate of the total treatment effectis expected to be
different under each design when interference is present.

Assume a network over the experimental units, such that the
immediate neighborhood $\N_i$ of unit $i$ are the units likely to interfere with
it, and the following model of potential outcomes:
\begin{equation}
  \label{eq:dirg}
  Y_i(\Z) = \alpha + \beta Z_i + \gamma \rho_i + \epsilon_i \quad (i =
  1,\ldots,N),
\end{equation}
where $\rho_i = \frac{1}{|\N_i|} \sum_{j \in \N_i} Z_j$ is
the average number of treated neighbors in unit $i$'s neighborhood $\N_i$ and
$\epsilon_i \sim \N(0, \sigma^2)$ is a noise factor, with $\epsilon_i \indep
\rho_i$. Interference is present if and only if $\gamma \neq 0$.
Hence, $\beta$ is often interpreted as a direct treatment effect
parameter, while $\gamma$ is often interpreted as an interference effect
parameter. Under this parametrized model of potential outcomes with
interference, the total treatment effect is $\tau = \beta + \gamma$.
\begin{lemma}\label{lem:cr_and_cbr_expectation}
  Under the model of interference in (\ref{eq:dirg}), the expectations of $\hat
  \tau_{cr}$ and $\hat \tau_{cbr}$ under their respective randomized designs
  are
\begin{equation*}
  \E_{\Z, \epsilon} ( \hat \tau_{cr} ) = \beta - \gamma (N - 1)^{-1}, \quad
  \E_{\Z, \epsilon} ( \hat \tau_{cbr} ) = \beta + \gamma (\rho_C
  M - 1)(M-1)^{-1}.
\end{equation*}
  where $\rho_C \defeq \frac{1}{N} \sum_{i=1}^N \frac{|\N_i \cap C(i)|}{|\N_i|}$
  is the average number of each unit's neighbors also present in its cluster,
  with $\calC(i)$ being the units belonging to unit $i$'s cluster.
\end{lemma}

Lemma~\ref{lem:cr_and_cbr_expectation} states that when interference is present,
neither estimator is unbiased for the total treatment effect, and, crucially,
they do not have the same expected value.

\subsection{A hierarchical randomization strategy}\label{sec:hier}\label{sec:intuition}
If it were possible to apply both the completely randomized and cluster-based
randomized designs to all experimental units, testing for
interference could be done by comparing the two estimates from each assignment strategy. If
the two estimates were significantly different, then there is evidence
interference is present. If the two estimates were not significantly different, we would
expect there to be no interference.

Unfortunately, just as both treatment and control cannot be assigned to each unit,
both assignment designs cannot be given to all experimental units. To solve this
problem, randomly assign units to treatment arms
and within each treatment arm, apply one experimental strategy for assigning units
to either intervention. In order to maintain some of the interference
structure intact within each treatment arm without sacrificing covariate balance
or introducing bias, we suggest to use a cluster-based randomized design to
assign units to treatment arms. Once clusters of units are assigned to one of
two treatment arms, we suggest to apply within each treatment arm either a
cluster-based randomized design or a completely randomized design.

The experimental units are grouped into $M$ balanced clusters, such that
each cluster has the same number of units.  Let $m_{cr}$ be the number of
clusters to be assigned to treatment arm $cr$ and $m_{cbr}$ be the number of
clusters to be assigned to treatment arm $cbr$. Let $n_{cr}$ and $n_{cbr} = N
- n_{cr}$ be the resulting number of units assigned to each arm.  Let $\omega
\in (0,1)^M$ be the cluster-to-treatment-arm assignment vector and $\W \in
(0,1)^N$ be the corresponding unit-to-treatment-arm assignment vector: $W_i =
1$ if unit $i$ is assigned to treatment arm $cr$ and $W_i = 0$ if unit $i$ is
assigned to treatment arm $cbr$.

In treatment arm $cr$, let $n_{cr,t}$ be the number of units that to be
assigned to treatment and $n_{cr,c}$ be the number of unit to be assigned
to control. Similarly, in treatment arm $cbr$, let $m_{cbr,t}$ and $m_{cbr,c}$
be the number of clusters that are assigned to treatment and control
respectively. Let $\Z \in (0, 1)^N$ be the assignment vector of units to
treatment and control, composed of two parts $\Z_{cr} \in (0, 1)^{n_{cr}}$ for
units in treatment arm $cr$ and $\Z_{cbr} \in (0,1)^{n_{cbr}}$ for units in
treatment arm $cbr$.

The hierarchical design is as follows: sample $\W$ in a cluster-based
randomized way. Conditioned on $\W$, sample $\Z_{cr}$ using a
completely randomized assignment to assign units in treatment arm $cr$ to
treatment and control. Conditioned on $\W$, sample $\Z_{cbr}$ using a
cluster-based randomized assignment to assign units in treatment arm $cbr$ to
treatment and control. The resulting assignment vector $\Z$ of units to
treatment and control is such that $\Z_{cr} \indep \Z_{cbr} | \W$.

Though the graph could be re-clustered for step $(iii)$, a simpler
option from an analytical and methodological perspective is to re-use the same
clustering used in step $(i)$. The two estimates of the causal effect
for this experiment as well as the difference-in-differences estimator
$\Delta$ are defined as follows:
\begin{align*}
  \hat \tau_{cr} & \defeq \overline{\Y}_{cr,t} - \overline{\Y}_{cr,c}, \\
  \hat \tau_{cbr} & \defeq \frac{m_{cbr}}{n_{cbr}} \left(
  \overline{\Y^+}_{cbr,t} - \overline{\Y^+}_{cbr,c}\right), \\
  \Delta &  \defeq \hat \tau_{cr} - \hat \tau_{cbr}.
\end{align*}
where $\Y_{cr,t}  \defeq (Y_i : W_i = 1 \wedge Z_i = 1)$ is the vector of
outcomes of units in treatment arm $cr$ that are treated.  Similarly, $\Y_{cr,c}
\defeq (Y_i : W_i = 1 \wedge Z_i = 0)$, $\Y^+_{cbr,t} \defeq (Y^+_j : \omega_j =
0 \wedge z_j = 1)$, $\Y^+_{cbr,c} \defeq (Y^+_j : \omega_j = 0 \wedge z_j = 0)$.
In the spirit of the Durbin--Hausman--Wu test, we could have chosen to compare
different estimators or different designs altogether, a comparison which is left to future work.

\section{Theory}\label{sec:theory}

\subsection{The Type I error}

In this section, we consider the expectation and the variance of the $\Delta$
estimator under the assumption of no interference in order to construct a
statistical test for whether interference is present. To simplify the analysis,
$n_{cr}$, $n_{cbr}$, $n_{cr,t}$, $n_{cr,c}$, $m_{cbr,t}$ and $m_{cbr,c}$ must be
constants, which implies that the clustering of the graph must be balanced. In
other words,  for any cluster $\calC_j \in \calC,~|\calC_j| = N/M =
n_{cr}/m_{cr} = n_{cbr}/m_{cbr}$.  Recall that $S_{tc}^+ = \sigma^2\{\Y^+(1) -
\Y^+(0)\}$ is the variance of the differences of the cluster-level outcomes,
that $\sigma^2_{cr}$ is the variance of the difference-in-means estimator under
a completely randomized assignment, and $\sigma^2_{cbr}$ is the variance of the
Horvitz--Thompson estimator under a cluster-based randomized assignment.
\begin{theorem}\label{thm:theor_var}
  If the stable unit treatment value assumption holds, and every cluster is the
  same size, then the expectation and variance of the difference-in-differences
  estimator are
 \begin{align*}
   \E_{\W,\Z}(\Delta) & = 0, \\ \var_{\W, \Z}(\Delta) & =  \sigma^2_{cr}  +
   \sigma^2_{cbr} + \frac{M}{n_{cr} n_{cbr}} S_{tc}^+ + O\left(\frac{M^2}{n_{cr}
   N^2}\sigma^2_{cr}\right).
 \end{align*}
\end{theorem}
The following corollary is a direct application of Chebyshev's inequality.

\begin{corollary}
  \label{cor:reject_null}
  Let the null hypothesis be that the stable unit treatment value assumption
  holds and let $\hat \sigma^2 \in \mathbbm{R}_+$ be any computable quantity
  from the experimental data which upper-bounds the true variance: $\hat
  \sigma^2 \geq \var_{\W, \Z}(\Delta)$.  Suppose that we reject the null if and
  only if $|\Delta| \geq \alpha^{-1/2} \sqrt{\hat \sigma^2}$, then if the null
  hypothesis holds, we reject the null (incorrectly) with probability no greater
  than $\alpha$.
\end{corollary}
This result holds for any balanced
clustering and for any model of interference because the
type I error assumes the null hypothesis. Another
way of rejecting the null is to approximate the test statistic $T \defeq (\hat
\mu_{cr} - \hat \mu_{cbr})(\hat \sigma^2)^{-1/2}$ by a normal distribution.  In
this case, a conservative $(1-\alpha) \times 100 \%$
confidence interval is $\left(T -z_{\alpha/2} , T +
z_{1-\alpha/2} \right)$, where $z_{\alpha/2}$ and $z_{1-\alpha/2}$ are the
$\alpha/2$-quantiles of the standard normal distribution.

\subsection{Variance estimators}\label{sec:variance}

Corollary~\ref{cor:reject_null} makes the assumption that $\hat \sigma^2$,
computable from observable data, is an upper-bound of the unknown theoretical
variance of the estimator $\sigma^2$. We discuss two solutions to finding the
smallest possible upper-bound $\hat \sigma^2$ of the theoretical variance.

The following variance estimator is inspired from Neymann's
conservative variance estimator, which upper-bounds the variance of the
difference-in-means estimator under a completely randomized assignment in
expectation. Consider the following empirical variance quantities in each
treatment bucket of each treatment arm:
$ \hat  S_{cr,t} \defeq \sigma^2(Y_i : W_i = 1 \wedge Z_i = 1)$ is the variance
of the observed outcomes of the treated units in the completely randomized
treatment arm, $ \hat  S_{cr,c} \defeq \sigma^2(Y_i : W_i = 1 \wedge Z_i = 0)$
is the variance of the observed outcomes of the control units in the completely
randomized treatment arm. Similarly, let $ \hat  S_{cbr,t}^+ \defeq
\sigma^2(Y_j^+ : \omega_j = 0 \wedge z_j = 1)$, and $ \hat  S_{cbr,c}^+ \defeq
\sigma^2(Y_j^+ : \omega_j = 0 \wedge z_j = 0)$ in the cluster-based randomized
arm.

\begin{theorem}\label{thm:var_ub}
Let $\hat \sigma^2$ be the variance estimator defined by
\begin{equation}\label{eq:var_ub}
  \hat \sigma^2 \defeq \frac{\hat S_{cr,t}}{n_{cr,t}} + \frac{\hat
  S_{cr,c}}{n_{cr,c}}+ \frac{m_{cbr}^2}{n_{cbr}^2} \left(\frac{\hat
  S_{cbr,t}^+}{m_{cbr,t}} + \frac{\hat S_{cbr,c}^+}{m_{cbr,c}} \right).
\end{equation}
  If the null hypothesis holds, then $\hat \sigma^2$ upper-bounds the
  theoretical variance of the difference-in-differences estimator $\Delta$ in
  expectation: $\E_{\W,\Z} \left(\hat \sigma^2 \right) \geq \var_{\W,
  \Z}(\Delta)$. Furthermore, in the case of a constant treatment effect, there
  exists $\tau \in \mathbbm{R}$ such that for all $i$, $Y_i(1) = Y_i(0) + \tau$,
  the inequality becomes tight: $\E_{\W,\Z} \left( \hat \sigma^2  \right) =
  \var_{\W, \Z}(\Delta)$.
\end{theorem}
The condition of Corollary~\ref{cor:reject_null} will be met only in
expectation. This is often deemed sufficient in the literature
\citep{imbens2015causal}. Due to its simplicity, we use this empirical
upper-bound in the analysis of our experiments on LinkedIn.

Another common upper-bound is obtained by assuming Fisher's null hypothesis of
no treatment effect, which posits that $Y_i(1) = Y_i(0)$ for all units $i$.  In
particular, this implies that there is no interference. The converse is not
true~\citep{rosenbaum2007interference}.  Under Fisher's null hypothesis, the
theoretical formula for the variance is computable from the observed data. Let
$S \defeq \sigma^2(\Y)$ be the variance of all observed outcomes, and $S^+
\defeq \sigma^2(\Y^+)$ be the variance of all observed aggregated outcomes.
\begin{theorem}\label{thm:fisher_null_var}
Under the null hypothesis of no treatment effect, if all clusters are the same size,
\begin{equation*}
  \var_{\W,\Z}(\Delta) = \frac{n_{cr}}{n_{cr} -1} \frac{M}{M-1}
  \frac{n_{cr}}{n_{cr,t} n_{cr,c}} S + \left\{1 - \frac{m_{cbr}}{N(n_{cr} -
  1)}\right\} \frac{m_{cbr}}{m_{cbr,t} m_{cbr,c}} S^+.
\end{equation*}
\end{theorem}


\subsection{The type II error}\label{sec:type_II}

To paraphrase the result stated in Corollary~\ref{cor:reject_null}, if the
rejection region is $\{ T \geq \alpha^{-1/2}\}$ and $\hat \sigma^2 \geq
\var_{W,Z}(\Delta)$, then the probability of falsely rejecting the null is lower
than $\alpha$. Computing the type I error is straightforward because the stable
unit treatment value assumption that the  holds.  The same is not true of the
type II error rate, where a model for the interference between units must be
assumed.

Section~\ref{sec:candidate_assign} looks at the expectation of both the
$\hat \tau_{cr}$ and $\hat \tau_{cbr}$ estimators for a completely randomized
and a cluster-based randomized assignments respectively under a
linear model of interference in (\ref{eq:dirg}).  We complete this analysis
by giving the type II error of the test under this same model of interference.
Recall that $\rho_C \defeq \frac{1}{N} \sum_{i=1}^N \frac{|\N_i \cap
\calC(i)|}{|\N_i|}$ is the average fraction of a unit's neighbors contained within
its cluster, as originally defined in Lemma~\ref{lem:cr_and_cbr_expectation}.
\begin{theorem}\label{thm:dirg}
  If all clusters are the same size, then under the linear model of
  interference defined in (\ref{eq:dirg}), the expectation  of the $\Delta$
  estimator under the suggested hierarchical design is
  $E_{\W, \Z} ( \Delta ) \approx \gamma \rho_C$
  under the assumption that $n_{cr} >> 1$, $m_{cbr}>>1$, and $m_{cr} >> 1$.
\end{theorem}
A proof is included in the supplementary material, which also includes the
computation of the variance of the $\Delta$ estimator under the above
interference model.  The result of Theorem~\ref{thm:dirg} is intuitive: when
$m_{cbr}$ and $n_{cr}$ are large, the stronger the interference (high
$|\gamma|$) and the better the clustering (high $\rho_C$), the larger the
expected difference between the two treatment arms.

Knowing the type II error rate can help determine which clustering of the
units is most appropriate. The selection of hyper-parameters in clustering
algorithms, including the number of clusters to set, can be informed by
minimizing the type II error under plausible models of interference.  The
optimization program $\max_{M, \calC} \rho_C (\hat \sigma^2_\calC)^{-1/2}$
depends on the choice of variance estimator $\hat
\sigma^2_\calC$ for a clustering $\calC$,
where $\calC$ is composed of $M$ balanced clusters. We discuss a reasonable
heuristic in Section~\ref{sec:cluster} to solving this optimization program,
conjectured to be NP-hard.


\section{Variations of the hierarchical design}
\label{sec:practical}

\subsection{Bernoulli randomization}\label{sec:bernoulli}
The completely randomized assignment is a well-understood assignment mechanism,
which avoids degenerate cases where all units are assigned to treatment and
control. However, experimentation platforms at major internet companies are
rarely set up to run completely randomized experiments. Instead, these platforms
run Bernoulli randomized assignments, which for large sample sizes, are
intuitively equivalent to completely randomized assignments.  We provide a
formal explanation for why running a Bernoulli randomized assignment does not
affect the validity of our test in practice: the variance of the
difference-in-means estimator under the Bernoulli randomized mechanism and the
completely randomized mechanism are equivalent up to $O(N^{-2})$ terms.
\begin{theorem}\label{thm:br_vs_cr}
  Let $CR$ be the completely randomized assignment, assigning exactly $n_t$
  units to treatment and $n_c \defeq N - n_t$ to control. Let $BR$ be the
  corresponding re-randomized Bernoulli assignment, assigning units to treatment
  with probability $p \defeq n_t/N$ and to control with probability $1 - p =
  n_c/N$. For all $N \geq 2$ such that $p^N + (1-p)^N \leq N^{-2}$, the
  following upper-bound holds
\begin{equation*}
  \left|\var_{\Z \sim BR}(\hat \tau) - \var_{\Z \sim CR}(\hat \tau) \right| \leq
  5 \left[\frac{\sigma^2\{\Y(1)\}}{n_t^2} + \frac{\sigma^2\{\Y(0)\}}{n_c^2} \right].
\end{equation*}
\end{theorem}
A proof, which considers a re-randomized Bernoulli assignment scheme that
rejects assignments where all units are assigned to treatment or to control, is
included in the supplementary material.

\subsection{Stratification and subsampling}\label{sec:strata}\label{sec:subsampling}

One practical concern with our suggested hierarchical design is that if the
chosen number of clusters is small, possibly much smaller than the number of
units, we run the risk of having strong covariate imbalances between the two
treatment arms. In this case, we recommend using a stratified treatment arm
assignment.  Let each graph cluster be assigned to one of $L$ strata.  Within
each stratum $s$, we assign clusters completely at random to treatment arm $cr$
and treatment arm $cbr$. Within each stratum $s$, units in treatment arm $cr$ are
assigned completely at random to treatment and control, while in treatment arm
$cbr$, clusters are assigned completely at random to treatment and control.  Let
$\hat \tau_{cr}(s)$, $\hat \tau_{cbr}(s)$, and $\Delta(s)$ be the restriction of
$\hat \tau_{cr}$, $\hat \tau_{cbr}$, and $\Delta$ respectively to stratum $s$.  Let
$M(s)$ be the total number of clusters in stratum $s$ and $M$ be the total
number of clusters. The stratified $\Delta'$ estimator and its empirical
variance upper-bound estimator can be expressed as a
weighted average of the $\Delta(s)$ and $\hat \sigma^2(s)$,
\begin{equation}
\label{eq:strat_delta}
   \Delta' \defeq \sum_{s =1}^L \frac{M(s)}{M} \Delta(s), \qquad
  \hat \sigma'^2 \defeq \sum_{s = 1}^L \left\{\frac{M(s)}{M} \right\}^2
  \hat \sigma^2(s),
\end{equation}
where $\hat \sigma^2(s)$ is the empirical upper-bound of $\var_{\W(s),
\Z(s)}\{\Delta(s)\}$ suggested in (\ref{eq:var_ub}).

An additional constraint for our suggested design is that online experimentation
platforms often need to run multiple experiments simultaneously, with multiple
values for treatment and control.  As a result, each experiment runs within a
segment of the population chosen completely at random, leaving the other units
available for other experiments. Since this subsampling might negatively impact
the quality of the clustering in the cluster-based randomized treatment arm, we
decided against subsampling at the unit level prior to the experiment. In other
words, we recommend subsampling at the cluster-level, prior to the assignment to
treatment arms, rather than at the unit level, when deciding which units to
include in the experiment.


\section{Simulation study}
\label{sec:simulation}

\begin{figure}
\centering
  \captionsetup{width=.85\linewidth}
\begin{tabular}{cc}
  \includegraphics[scale=.54]{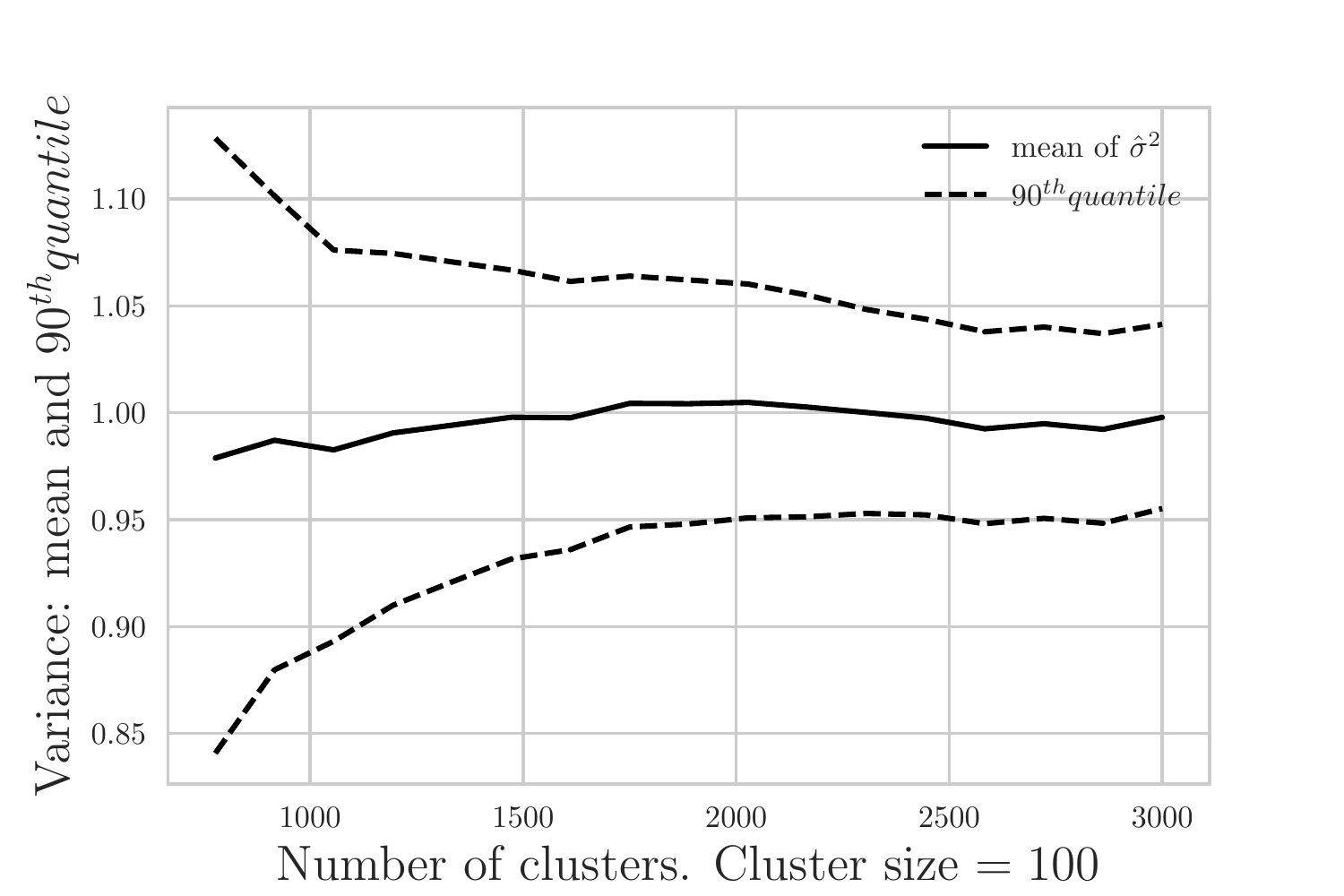} &
  \includegraphics[scale=.54]{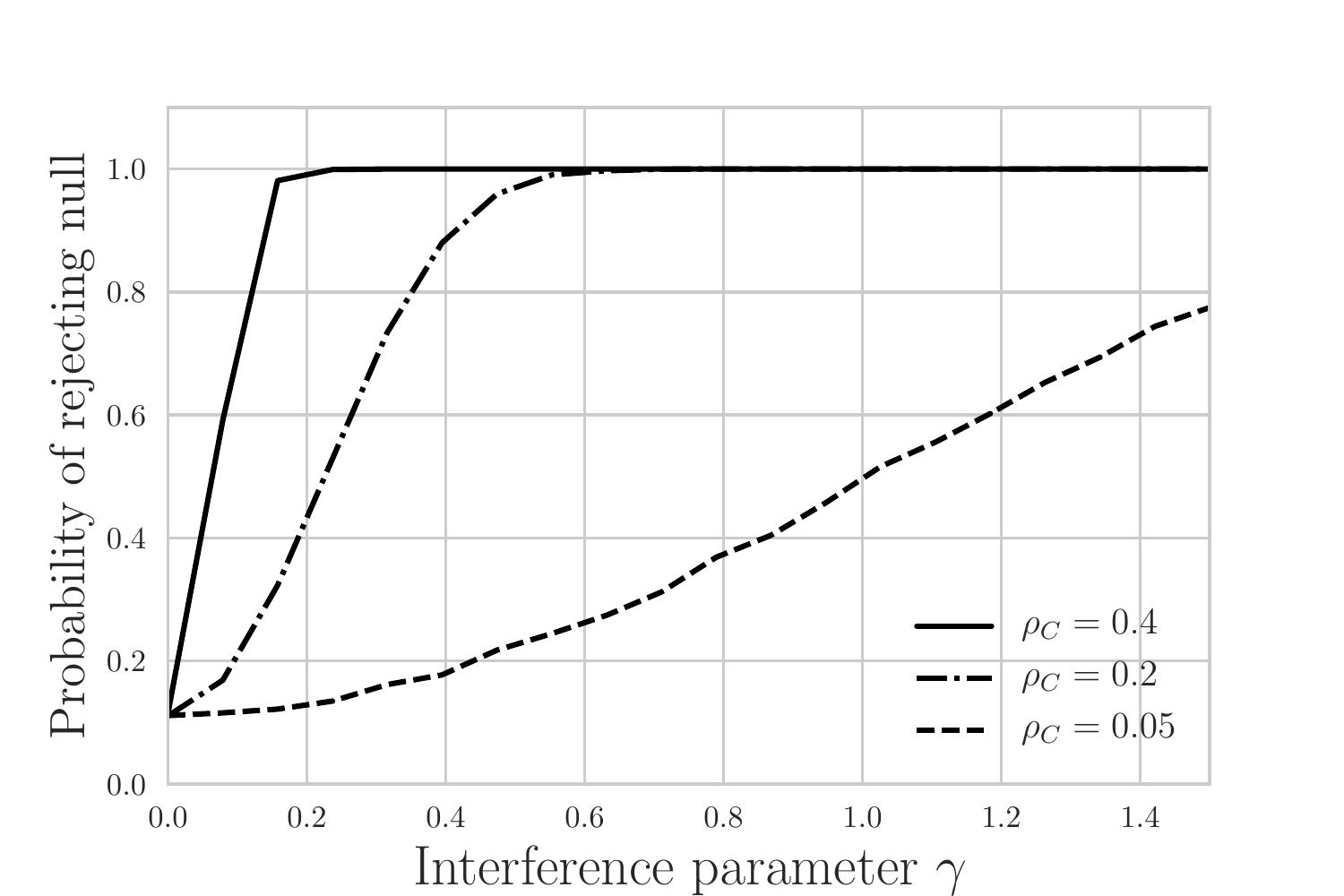} \\
  (a) & (b)
\end{tabular}
  \caption{\emph{(a)}~The expectation, $10^{th}$ and $90^{th}$
  quantile of the ratio of the empirical upper-bound estimator $\hat \sigma^2$
  over the true variance $\var_{\W, \Z}(\Delta)$. \emph{(b)}~Probability of
  rejecting the null of no interference, under the linear interference model in
  (\ref{eq:dirg}).}\label{fig:power}
\end{figure}

The effect of the
clustering on the type I and type II error of the test for interference can
better be understood through simulation.
From Theorem~\ref{thm:var_ub},
the bound on the type I error holds under the
assumption that the empirical upper-bound for the variance upper-bounds the
theoretical variance for the realized assignment $\hat \sigma^2(\Z) \geq
\var_\Z(\Delta)$, a property that is only guaranteed in
expectation. In a first simulation study, we examined how often it held in
practice for different numbers of clusters, from $500$ to $3000$, of fixed size,
$100$ units. Fig.~\ref{fig:power}.a reports the
expectation and the $10^{th}$ and $90^{th}$ quantiles of the ratio of the
empirical upper-bound $\hat \sigma^2$ over the true
variance $\var_{W, Z}(\Delta)$. For each point on the x-axis, $5 \times 10^5$
simulations were run. The upper-bound holds (tightly) in expectation
but is not an upper-bound almost surely. Despite the diminishing
returns on the reduction of the confidence intervals from increasing the number
of clusters, for a number of cluster greater than $2000$, the ratio is in the
$(.95,~1.05)$ range more than $90\%$ of the time.

In a second simulation study, we quantified the type II error of our test under
thelinear interference model in
(\ref{eq:dirg}), fixing the value of the constant parameter to $0$
and the direct treatment effect parameter to $1$, and choosing as a graph a block-model with $40$
balanced clusters of $1000$ units each. The probability of an edge existing
between two units of the graph is a constant cluster-level probability, set
consecutively to
$(.01,~.31)$, $(.15,~.45)$, and $(.3,~.6)$, denoting the
intra-cluster and inter-cluster probabilities. These three tuples correspond to
values of the graph cut parameter
$\rho_C$ equal to $(.05,~.2,~.4)$ respectively, defined in
Lemma~\ref{lem:cr_and_cbr_expectation}. The higher $\rho_C$, the fewer edges of
the graph are cut. The value of the interference parameter was varied from $0$
to $1.4$.  Fig.~\ref{fig:power}.b reports the probability of rejecting the null
under $1000$ simulations. Even with a low value of the graph cut parameter,
$\rho_C = .05$, the test of interference correctly rejects the null under levels
of interference that are of similar magnitude to the direct effect $75\%$ of the
time. Furthermore, if $\rho_\calC \geq 0.4$, the test correctly rejects the null
for levels of interference that are at least $1/5$th of the magnitude of the
direct treatment effect $99.9\%$ of the time.


\section{Illustration on a LinkedIn experiment}\label{sec:results}

\subsection{Experimental set-up}

Google~\citep{tang2010overlapping},
Microsoft, \citep{kohavi2013online}, Facebook~\citep{bakshy2014designing},
LinkedIn~\citep{xu2015infrastructure}, and other major technology companies rely
on experimentation to understand the effect of each product decision, from minor
user interface changes to major product launches. Due to their extensive
reliance on randomized experiments, each company has built mature
experimentation platforms.  It is an open question how many of these experiments
suffer from interference effects. By collaborating with the team in charge of
LinkedIn's experimentation platform, we tested for interference in several of
LinkedIn's many randomized experiments.

Users on LinkedIn can interact with content posted by their neighbors through a
personalized feed. Rather than presenting the content chronologically, LinkedIn
strives to improve a user's feed by ranking content by relevance. In order to
improve user experience, researchers at LinkedIn suggest new feed ranking
algorithms and seek to determine the impact of each algorithm on user metrics
through a randomized experiment. These metrics may include time spent on the
site, engagement with content on the feed, and original content creation.
Experimentation on feed ranking algorithms is a typical case where interference
between units is present. If a user is assigned to a better feed ranking
algorithm, they will interact more with their feed by liking or sharing content
more. This in turn impacts what her connections see on their own feed. We seek
to understand whether or not these network effects are negligible.

To run the experiment, we clustered the LinkedIn graph into balanced clusters (Section~\ref{sec:cluster}), stratified the clusters by blocking on cluster covariates (Section~\ref{sec:strata_practical}),
assigned a subset of clusters to treatment and to control chosen at random, comprising the second treatment arm and treatment bucket assignment.
Any unit not already assigned to treatment or control was given to the main
experimentation pipeline. A sub-population of units is first sampled at
random (Section~\ref{sec:subsampling}) and then assigned to treatment
and control using a Bernoulli randomized assignment (Section~\ref{sec:bernoulli}).

Before applying treatment to units assigned to treatment, we ran covariate
balance checks and measured outcomes $4$ and $2$ months prior to the launch of
the experiment, on the day of the launch, and again $2$ months after the launch.
The number of units per treatment arm  was in the order of
several million.

\subsection{Graph clustering}\label{sec:cluster}
The main challenge of implementing the proposed test for interference is
clustering the graph into clusters of equal size.  Only parallelizable
algorithms can operate at the scale of the
LinkedIn graph.  \citet{kdd17} performed an extensive experimental evaluation of the state-of-the-art balanced
clustering algorithms and found the restreaming version of the Linear
Deterministic Greedy algorithm to work
best~\citep{nishimura2013restreaming}. We ran the parallel version of this
algorithm for $30$ iterations, set the number of clusters to
$M=3000$ and a leniency of $1\%$ for the balance
constraint, to slightly sacrifice balance for better clustering quality, as it
compromised between maximizing the fraction of edges within clusters ($28.28\%$)
and minimizing pre-treatment variance.

\subsection{Clusters stratification}\label{sec:strata_practical}
As suggested in Section~\ref{sec:strata}, each cluster is assigned to one of $L$
strata in order to reduce the variance of the estimator and to ensure the balance of
cluster-level covariates. Each cluster is described by the number
of edges within the cluster, the number of edges with an endpoint in another
cluster, and two metrics that characterize users' engagement with the LinkedIn
feed (averaged over all users in the cluster). \citet{kdd17} found that stratification using balanced
k-means clustering to work best~\citep{malinen2014balanced}.

\subsection{Results}
We launched our experimental design on LinkedIn in August 2016. Because of the
nature of our intervention on the social nature of the LinkedIn feed, we
expected the experiment to suffer from strong interference effects.  The primary
outcome was the change in a user's engagement over time, $Y_i(t) = y_i(t) -
y_i(t - 2)$, where $t - 2$ takes place two months before date $t$. As a sanity
check, we ran an A/A test on $Y_i(-2) = y_i( - 2) - y_i(- 4)$, where $t = 0$ is
the month the intervention was launched, $t = -2$ takes place two months prior,
and $t = -4$ takes place four months prior.  As expected, we found that no
significant interference in the A/A test, with a p-value of $0.68$ using the
gaussian assumption from Corollary~\ref{cor:reject_null}. We then evaluated the
presence of interference two months after the launch of the randomized
experiment, finding a p-value of $0.048$, and concluding that interference was
present in the experiment, as reported in Table~\ref{tab:main_results}.
Outcomes have been multiplied by a constant to avoid disclosing raw numbers.
$\Delta'$ and $\sigma'$ are defined in (\ref{eq:strat_delta}).

\begin{table}
  \centering
\begin{tabular}{ccc}
  Statistic  & Pre-treatment & Post-treatment  \\
  $\Delta'$ & $-3.3$ &  $-16.4$ \\
  $\sigma'^2$ & $8.1$ & $8.5$ \\
  2-tailed p-value   & $0.68$ & $0.048$ \\
\end{tabular}
\caption{Results of experiments on a subset of the LinkedIn
  graph}\label{tab:main_results}
\end{table}

\section*{Acknowledgement}

Guillaume Saint-Jacques and Martin Saveski contributed equally to this work,
while interning at LinkedIn, Sunnyvale, California. The authors wish to thank
Guillaume Basse and Dean Eckles for useful comments and discussions.  This
research was sponsored, in part, by National science Foundation awards CAREER
IIS-1149662 and IIS-1409177, and by Office of Naval Research awards YIP
N00014-14-1-0485 and N00014-17-1-2131.

\bibliography{paper}
\bibliographystyle{apalike}

\newpage

\appendix
\section{Proofs and supplementary material}


Supplementary material includes the proofs of
Lemma~\ref{lem:cr_and_cbr_expectation},
Theorems~\ref{thm:theor_var}--\ref{thm:br_vs_cr}, and
Corollary~\ref{cor:reject_null}.

For any vector $u \in \mathbbm{R}^N$, let $\bar u = N^{-1} \sum_{i
=1}^N u_i$ and $\sigma^2(u) = (N-1)^{-1} \sum_{i = 1}^N (u_i - \bar u)^2$.

\section*{Proof of Lemma~\ref{lem:cr_and_cbr_expectation}}
For each unit $i$, let $\rho_i \defeq |\N_i|^{-1} \sum_{j \in \N_i} Z_j$ and the
potential outcomes model is
\begin{equation*}
  Y_i(\Z) = \alpha + \beta Z_i + \gamma \rho_i + \epsilon_i \quad (i = 1,\dots,
  N).
\end{equation*}
The expectation of the difference-in-means estimator under a completely
randomized assignment is
\begin{align*}
  \E_{\Z \sim cr} \left(\hat \tau_{cr} \right) & = \E_{\Z\sim cr} \left\{ \sum_{i=1}^N \frac{(-1)^{Z_i}}{n_t^{Z_i} n_c^{(1- Z_i)}} \left(\alpha + \beta Z_i +
  \frac{\gamma}{|\N_i|} \sum_{j \in \N_i} Z_j \right) + \epsilon_i \right\} = \beta - \frac{\gamma}{N-1}.
\end{align*}
The expectation of the Horvitz-Thompson estimator under a cluster-based
randomized assignment:
\begin{align*}
  \E_{\Z \sim cr}(\hat \tau_{cbr}) &= \E_{\Z \sim cr} \left\{\frac{M}{N} \sum_{i = 1}^N
  \frac{(-1)^{Z_i}}{m_t^{Z_i} m_c^{(1- Z_i)}} \left(\alpha + \beta Z_i +
  \frac{\gamma}{|\N_i|} \sum_{j \in \N_i} Z_j \right) \right\} = \beta - \gamma \frac{1- \rho_C \cdot M}{M-1},
\end{align*}
where $\rho_C = N^{-1} \sum_{i=1}^N  |\N_i \cap
C(i)|/|\N_i|$.

\section*{Proof of Theorem~\ref{thm:theor_var}}

Let $m_{cr}$ and $n_{cr}$ be the number of clusters and units respectively in
treatment arm $cr$, and let $m_{cbr}$ and $n_{cbr}$ be the number of clusters
and units respectively in treatment arm $cbr$, such that $M = m_{cr} + m_{cbr}$
and $N = n_{cr} + n_{cbr}$.  Assume that the clustering of the graph is
balanced: $M/N =  m_{cr}/n_{cr} = m_{cbr}/n_{cbr}$.  If the stable unit
treatment value assumption holds, then $Y_i(\Z) = Y_i(Z_i)$. Under this
assumption,
\begin{align*}
  \E_{\Z}(\Delta | \W) & = \frac{1}{n_{cr}} \sum_{i =
  1}^N W_i\{Y_i(1) - Y_i(0)\} - \frac{1}{n_{cbr}} \sum_{i = 1}^N (1 - W_i) \{Y_i(1)
  - Y_i(0)\}. \\
  \E_{\W, \Z} (\Delta) & = \frac{1}{N} \sum_{i = 1}^N
  Y_i(1) - Y_i(0) - \frac{1}{N} \sum_{i = 1}^N Y_i(1) - Y_i(0) = 0.
\end{align*}
By Eve's law, the theoretical variance of the $\Delta$ estimator is
\begin{align*}
  \var_{\W,\Z}(\Delta) = \E_\W\{\var_\Z(\hat \tau_{cr} | \W)\}  + \E_\W
  \{\var_\Z(\hat \tau_{cbr}| \W)\} + \var_\W \{\E_\Z(\hat \tau_{cr} - \hat
  \tau_{cbr} | \W)\}
\end{align*}

Each term can be computed separately. The number of treated and control units in
treatment arm $cr$ are $n_{cr,t}$ and $n_{cr,c}$ respectively; $n_{cbr,t}$ and
$n_{cbr,c}$ are the number of treated and control units in treatment arm $cbr$.
The following quantities are variances of potential outcomes, restricted to the
treatment arms $cr$ and $cbr$, and thus expressed conditionally on the
assignment $\W$ of clusters to each treatment arm.  Let $S_{cr, t} =
\sigma^2\{Y_i(1) : W_i = 1\}$, $S_{cr, c} = \sigma^2\{Y_i(0) :  W_i = 1\}$ and
$S_{cr, tc} \defeq \sigma^2\{Y_i(1) - Y_i(0) : W_i = 1\}$.  Let $\omega \in
(0,1)^M$ be the cluster indicator vector for whether a cluster has been
assigned to treatment arm $cr$ or to
$cbr$: $\omega$ is a cluster-level representation $\W$. Let $S^+_{cbr, t} \defeq
\sigma^2\{Y^+_j(1) : \omega_j = 0)$, $S^+_{cbr, c} \defeq \sigma^2(Y^+_j(0) :
\omega_j = 0\}$ and $S_{cbr, tc} \defeq \sigma^2\{Y^+_j(1) - Y^+_j(0) : \omega_j =
0\}$. Conditioned on the assignment of units to treatment arms,
\begin{align*}
   \var_{\Z \sim cr}(\hat \tau_{cr} | W) = \frac{S_{cr,t}}{n_{cr,t}}  +
  \frac{S_{cr,c}}{n_{cr,c}}  - \frac{S_{cr, tc}}{n_{cr}}, \quad
  \var_{\Z \sim cbr}(\hat \tau_{cbr} | W)  = \frac{m_{cbr}}{n_{cbr}} \left(
  \frac{S_{cbr,t}}{m_{cbr,t}}  + \frac{S_{cbr,c}}{m_{cbr,c}}  -
  \frac{S_{cbr, tc}}{m_{cbr}} \right).
\end{align*}
By linearity of expectation, we compute the expectation of each term with
respect to $\W$, beginning with the cluster-based randomized treatment arm.
\begin{align*}
  \E_\W(S_{cbr, t}) = \frac{m_{cbr}}{m_{cbr} - 1} \overline{\Y^+(1)^2} -
  \frac{m_{cbr}}{m_{cbr} -1} \E \left(\overline{\Y^+_{cbr}(1)}^2 \right).
\end{align*}
Let $d_j = \omega_j - \frac{m_{cbr}}{m}$. The second term is
\begin{align*}
  \E_\W \left\{ \overline{\Y^+_{cbr}(1)}^2 \right\}
  &= \frac{1}{m_{cbr}^2} \sum_{j \in \calC} \sum_{k \in \calC} \E[ \omega_j \omega_k]
  Y^+_j(1) Y^+_k(1) \\
  &= \frac{1}{m_{cbr}^2} \sum_{j \in \calC}  \E \left[ d_j^2 + \frac{m_{cbr}^2}{M^2}\right]
  (Y^+_j(1))^2 + \frac{1}{m_{cbr}^2} \sum_{j \in \calC} \sum_{k \neq j} \E \left[ d_j d_k
  + \frac{m_{cbr}^2}{M^2}\right] Y^+_j(1) Y^+_k(1) \\
  &= \frac{1}{m_{cbr}^2} \left(\frac{m_{cbr} (M -m_{cbr})}{M^2} +
\frac{m_{cbr}^2}{M^2} \right) \cdot M \cdot \overline{(Y^+(1))^2}
 \\ & \quad + \frac{1}{m_{cbr}^2} \left(- \frac{m_{cbr} (M - m_{cbr})}{M^2 (M-1)} +
\frac{m_{cbr}^2}{M^2}\right) \sum_{j \in \calC} \sum_{k \neq j} Y_j(1) Y_k(1) \\
  &= \frac{1}{m_{cbr}} \cdot \overline{(Y^+(1))^2} +
  \frac{1}{m_{cbr}^2}\frac{m_{cbr}}{M^2}\frac{M (m_{cbr}-1)}{M-1} \sum_{j \in
  \calC} \sum_{k
  \neq j} Y_j(1) Y_k(1) \\
  &= \left(\frac{1}{m_{cbr}} - M\cdot \frac{m_{cbr}-1}{m_{cbr} M (M-1)}\right)
  \overline{(Y^+(1))^2} + \frac{m_{cbr} -1}{m_{cbr} M (M-1)} \sum_{j \in \calC}
  \sum_{k \in \calC}
Y^+_j(1) Y^+_k(1) \\
&= \frac{m_{cr}}{m_{cbr} (M-1)}\overline{(Y^+(1))^2} + \frac{M(m_{cbr}-1)}{m_{cbr} (M-1)} \left(
\overline{Y^+(1)} \right)^2
\end{align*}
Putting both terms together, $\E_W(S_{cbr,t}) = \sigma^2\{\Y^+(1)\} = S^+_t$.
Similarly, $\E_W(S_{cbr,c}) = S^+_c$ and $\E_W(S_{cbr, tc}) = S^+_{tc}$. Therefore,
\begin{equation*}
  \E_\W \{ \var(\hat \tau_{cbr} |\W) \} = \frac{m_{cbr}^2}{n_{cbr}^2} \left(
  \frac{S^+_t}{m_{cbr,t}} + \frac{S^+_c}{m_{cbr,c}} - \frac{S^+_{tc}}{m_{cbr}}
  \right).
\end{equation*}
Repeating the analysis for the treatment arm $cr$,
\begin{align*}
  \E_\W(S_{cr,t})
  &= \frac{1}{n_{cr} - 1} \sum_{i =1}^N \E_\W(W_i) Y_i^2(1) -
  \frac{n_{cr}}{n_{cr} - 1} \E_\W\left\{\overline{\Y_{cr}(1)}^2
  \right\}.
\end{align*}
The first term is $\sum_{i=1}^N \E_\W(W_i) Y_i^2(1) =
n_{cr}\overline{\Y^2(1)}$. Noting that $\overline{\Y_{cr}(1)} = m_{cr}
n_{cr}^{-1} \overline{\Y^+_{cr}(1)}$ and $\overline{\Y(1)} = m_{cr}n_{cr}^{-1}
\overline{\Y^+(1)}$, and by analogy from the computation for the treatment arm
$cbr$, the second term is
\begin{align*}
  \E_\W \left( \overline{\Y_{cr}(1)}^2 \right) & = 
 \frac{m_{cr}^2}{n_{cr}^2} \left( \frac{m_{cbr}}{m_{cr} (M-1)} \overline{(Y^+)^2(1)} +
\frac{M (m_{cr} - 1)}{m_{cr} (M-1)} \left(\overline{Y^+(1)} \right)^2\right) \\
&= \frac{m_{cr}^2}{n_{cr}^2} \left( \frac{m_{cbr}}{m_{cr} (M-1)} \overline{(Y^+)^2(1)} + \frac{M
(m_{cr} - 1)}{m_{cr} (M-1)} \frac{n_{cr}^2}{m_{cr}^2} \left(\overline{Y(1)} \right)^2\right)
\\
  &  = \frac{1}{n_{cr}^2}
  \frac{m_{cr} m_{cbr}}{M -1} \overline{(\Y^+)^2(1)} + \frac{N}{n_{cr}}
  \frac{m_{cr} - 1}{M -1} \overline{\Y(1)}^2.
\end{align*}
Putting the two terms together and letting $S_t \defeq \sigma^2\{\Y(1)\}$ and $S_t^+ \defeq
\sigma^2\{\Y^+(1)\}$,
\begin{align*}
  \E_\W(S_{cr, t}) &=  \frac{n_{cr}}{n_{cr} - 1} \frac{M-1}{M} S_t -
  \frac{m_{cbr}}{N (n_{cr} - 1)} S_t^+.
\end{align*}
Similarly,
\begin{equation*}
  \E_\W(S_{cr,c}) = \frac{n_{cr}}{n_{cr} - 1} \frac{M-1}{M} S_c
   - \frac{m_{cbr}}{N (n_{cr} - 1)} S_c^+, \quad 
  \E_\W(S_{cr,tc}) = \frac{n_{cr}}{n_{cr} - 1} \frac{M-1}{M} S_{tc} -
  \frac{m_{cbr}}{N (n_{cr} - 1)} S_{tc}^+.
\end{equation*}
where $S_c \defeq \sigma^2(\Y(0))$ and $S_{tc} \defeq
\sigma^2\{\Y(1) - \Y(0)\}$.
Using the approximation that $n_{cr} (M -1)\{M (n_{cr} - 1)\}^{-1} = 1 +
O\left(n_{cr}^{-1}\right)$, and the Cauchy-Schwarz inequality, $S^+_t = O\left(M
N^{-1} S_t \right)$,
\begin{align*}
  \E_\W\{\var(\hat \tau_{cr} | \W)\} & = \sigma_{cr}^2 + O\left( \frac{M^2}{n_{cr}
  N^2 } \sigma_{cr}^2\right).
\end{align*}
Finally, letting $S_{tc}^+ = \sigma^2\{\Y^+(1) - \Y^+(0)\}$,
\begin{align*}
  \var_\W\{\mathbbm{E}_\Z(\hat \tau_{cr} - \hat \tau_{cbr} |\W)\}
&=\var\left[ \frac{1}{n_{cr}} \sum_{i =1}^N W_i(Y_i(1) - Y_i(0)) -
  \frac{1}{n_{cbr}} \sum_{i = 1}^N (1 - W_i) (Y_i(1) - Y_i(0)) \right] \\
&= \var \left[ \frac{1}{N} \sum_{i =1 }^N \left(W_i - \frac{m_{cr}}{M} \right)
  \left(\frac{M}{m_{cr}} + \frac{M}{m_{cbr}} \right) (Y_i(1) - Y_i(0))\right]\\
  &=   \frac{M^2}{N^2}  \frac{m_{cr} m_{cbr}}{M^3 (M-1)} \frac{M^4}{m_{cr}^2
  m_{cbr}^2} \sum_{j = 1}^M \left[Y^+_j(1) - Y^+_j(0) -
  \left\{\overline{\Y^+(1)} - \overline{\Y^+(0)}\right\} \right]^2 \\
&= \frac{M}{n_{cr} n_{cbr}} S_{tc}^+.
\end{align*}


\section*{Proof of Corollary~\ref{cor:reject_null}}

Under the stable unit treatment value assumption, the test statistic $T \defeq
(\hat \tau_{br} - \hat \tau_{cbr})\hat \sigma^{-2}$ is a random variable with mean
$0$ and variance smaller than $1$ if $\hat \sigma^2 \geq \sigma^2$. Under this
assumption, from Chebyshev's inequality, if we reject with $\{ |T| \geq
\alpha^{-1/2}$, then we reject with probability less than $\alpha$.


\section*{Proof of Theorem~\ref{thm:var_ub}}

Let $S_{cr,t} \defeq \sigma^2(Y_i(1) : W_i = 1)$, $S_{cr,c} \defeq
\sigma^2(Y_i(0) : W_i = 1)$, $S_{cr,tc} \defeq \sigma^2(Y_i(1) - Y_i(0) : W_i
= 1)$,  $S^+_{cbr,t} \defeq \sigma^2(Y^+_j(1) : \omega_j =
0)$, $S^+_{cbr,c} \defeq \sigma^2(Y^+_j(0) : \omega_j = 0)$, and $S_{cbr,tc}
\defeq \sigma^2(Y_j(1) - Y_j(0) : \omega_j = 0)$. The theoretical variance is
\begin{align*}
  \var_{\W,\Z}(\Delta) &= \var_\W \left\{ \E_\Z( \hat \tau_{cr} - \hat
  \tau_{cbr}| \W)\right\} + \frac{\E_\W( S_{cr,t})}{n_{cr,t}}  +
  \frac{\E_\W( S_{cr,c})}{n_{cr,c}}  - \frac{\E_\W(S_{cr,tc})}{n_{cr}} \\
  & \quad +  \frac{m_{cbr}^2}{n_{cbr}^2} \left\{ \frac{\E_\W(S_{cbr,t}
  )}{m_{cbr,t}}   + \frac{\E_\W( S^+_{cbr,c})}{m_{cbr,c}}  - \frac{\E_\W ( S^+_{cbr,tc})}{m_{cbr}}
   \right\}.
\end{align*}
Since $\E_\W \left(S_{cr,tc}\right) \geq 0$ and $\E_\W \left( S^+_{cbr,tc}
\right) = S_{tc}^+$ and  $\var_\W \left\{ \E_\Z( \hat \tau_{cr} - \hat \tau_{cbr} | \W)\right\} =
  m (n_{cr} n_{cbr})^{-1} S_{tc}^+$, we observe that $\left\{ - m_{cbr}^2 n_{cbr}^{-2} m_{cbr}^{-1} + M (n_{cr}
n_{cbr})^{-1} \right\} S_{tc}^+ < 0$.  Thus,

\begin{equation*}
  \var_{\W,\Z}(\Delta) \leq \frac{1}{n_{cr,t}}
  \E_\W( S_{cr,t}) + \frac{1}{n_{cr,c}} \E_\W( S_{cr,c}) +
  \frac{m_{cbr}^2}{n_{cbr}^2}
  \left\{ \frac{1}{m_{cbr,t}} \E_\W( S^+_{cbr,t} )  + \frac{1}{m_{cbr,c}} \E_\W(
  S^+_{cbr,c} ) \right\}.
\end{equation*}

Let $ \hat  S_{cr,t} \defeq \sigma^2(Y_i : W_i
= 1 \wedge Z_i = 1)$, $ \hat  S_{cr,c} \defeq \sigma^2(Y_i : W_i = 1 \wedge Z_i
= 0)$, $ \hat  S_{cbr,t} \defeq \sigma^2(Y_i : W_i = 0 \wedge Z_i = 1)$, $ \hat
S_{cbr,c} \defeq \sigma^2(Y_i : W_i = 0 \wedge Z_i = 0)$. Consider the empirical quantity
\begin{equation*}
  \hat \sigma^2 \defeq \frac{\hat S_{cr,t}}{n_{cr,t}} + \frac{\hat
  S_{cr,c}}{n_{cr,c}}+ \frac{m_{cbr}^2}{n_{cbr}^2} \left(\frac{\hat
    S_{cbr,t}^+}{m_{cbr,t}} +
  \frac{\hat S_{cbr,c}^+}{m_{cbr,c}} \right).
\end{equation*}
The following equalities hold: $\E_\Z( \hat S_{cr,t} |\W) = S_{cr,t}$, $\E_Z(
  \hat S_{cr,c} |\W) = S_{cr,c}$, $\E_\Z( \hat S^+_{cbr,t} |\W) =
S^+_{cbr,t}$, $\E_Z( \hat S^+_{cbr,c} |\W) = S^+_{cbr,c}$. As a result,
$\E_{\W, \Z} \left( \hat \sigma^2 \right) \geq \var_{\W, \Z}(\Delta)$.


\section*{Proof of Theorem~\ref{thm:fisher_null_var}}

Under Fisher's null, for all $i$, $Y_i(1) = Y_i(0)$. By substitution,
\begin{equation*}
  \var_{\W,\Z}(\Delta) =  \frac{n_{cr}}{n_{cr} - 1}\frac{M}{M-1}
  \frac{n_{cr}}{n_{cr,t} n_{cr,c}} S + \left\{1 - \frac{m_{cbr}}{N(n_{cr}
  -1)}\right\} \frac{m_{cbr}}{m_{cbr,t}m_{cbr,c}} S^+,
\end{equation*}
where $S \defeq \sigma^2(\Y)$ is the variance of all observed potential
outcomes and $S^+ \defeq \sigma^2(\Y^+)$ is the variance of all observed
\emph{aggregated} outcomes.

\section*{Proof of Theorem~\ref{thm:dirg}}

We now compute the expectation of both $\hat \tau_{cr}$ and $\hat \tau_{cbr}$
under their respective completely-randomized and cluster-based randomized
assignment assuming the following model of interference:
\begin{equation}
  \label{eq:linear_interference}
  Y_i(\Z) = \alpha + \beta Z_i + \gamma \rho_i + \epsilon_i \quad (i = 1,\dots,
  N).
\end{equation}
$\Delta$ can be decomposed into three differences $\Delta = (\hat \tau_{cr, \alpha} - \hat \tau_{{cbr}, \alpha}) + (\hat
  \tau_{cr, \beta} - \hat \tau_{cbr, \beta}) + (\hat \tau_{cr,\gamma} - \hat
  \tau_{cbr, \gamma})$, where
\begin{align*}
  \hat \tau_{cr, \alpha}& \defeq \sum_{i=1}^N W_i (-1)^{1 - Z_i}
  \frac{1}{n_{cr,t}^{Z_i}} \frac{1}{n_{cr,c}^{1 - Z_i}} (\alpha + \epsilon_i),\\
  \hat \tau_{cr, \beta}& \defeq \beta \sum_{i=1}^N W_i (-1)^{1 - Z_i}
  \frac{1}{n_{cr,t}^{Z_i}} \frac{1}{n_{cr,c}^{1 - Z_i}} Z_i,\\
  \hat \tau_{cr, \gamma}& \defeq \gamma \sum_{i =1}^N W_i (-1)^{1 - Z_i}
  \frac{1}{n_{cr,t}^{Z_i}} \frac{1}{n_{cr,c}^{1 - Z_i}} \frac{1}{|\N_i|}
  \sum_{j \in \N_i} Z_j, \\
  \hat \tau_{cbr, \alpha}& \defeq \frac{m_{cbr}}{n_{cbr}} \sum_{i=1}^N (1 - W_i)
  (-1)^{1 - Z_i} \frac{1}{m_{cbr,t}^{Z_i}} \frac{1}{m_{cbr,c}^{1 - Z_i}} (\alpha
  + \epsilon_i),\\
  \hat \tau_{cbr, \beta}& \defeq  \beta \frac{m_{cbr}}{n_{cbr}}\sum_{i=1}^N (1 -
  W_i) (-1)^{1 - Z_i} \frac{1}{m_{cbr,t}^{Z_i}} \frac{1}{m_{cbr,c}^{1 - Z_i}}
  Z_i,\\
  \hat \tau_{cbr, \gamma}& \defeq \gamma \frac{m_{cbr}}{n_{cbr}}\sum_{i=1}^N (1
  - W_i) (-1)^{1 - Z_i} \frac{1}{m_{cbr,t}^{Z_i}} \frac{1}{m_{cbr,c}^{1 - Z_i}}
  \frac{1}{|\N_i|} \sum_{j \in \N_i} Z_j.
\end{align*}
The expectation of the first difference with respect to $(\W, \Z)$ is $0$ since
$\E_\Z( \hat \tau_{cr,\alpha} | \W )  = 0 = \E_\Z(\hat \tau_{cbr,\alpha} | \W)$.
Similarly, the second differene is also equal to $0$ since $\E_\W \left\{\E_\Z(
\hat \tau_{cr,\beta} | \W )\right\}  = \beta =  \E_\W \left\{ \E_\Z( \hat
\tau_{cbr,\beta} | \W ) \right\}$.  In order to simplify the calculus of the
third difference, suppose that $n_{cr} >> 1$ and $m_{cbr}$ such that $Z_i$ and
$Z_j$ can be considered independent if both $i$ and $j$ are in the same
treatment arm, except if they are in the cluster-based treatment arm and belong
to the same cluster. Under this assumption, $\E_\Z( \hat \tau_{cr,\gamma} | \W)
= 0$. Under the second arm,
\begin{align*}
  \E_\Z( \hat \tau_{cbr,\gamma} | \W) &= \gamma \frac{m_{cbr}}{n_{cbr}}\sum_{i=1}^N (1 - W_i) \frac{1}{|\N_i|}
  \sum_{j \in \N_i \cap C(i)} \frac{1- W_j}{m_{cbr}}.
\end{align*}
Taking the expectation over $\W$, $\E_{\W, \Z} (\hat\tau_{cbr, \gamma}) = \gamma
N^{-1} \sum_{i=1}^N |\N_i\cap C(i)| |\N_i|^{-1}$. Then, $\E_{\W, \Z} \left(\hat
\tau_{cr} - \hat \tau_{cbr} \right)  \approx \rho_C \cdot \gamma,$ where it is
assumed that $n_{cr} >> 1$, $m_{cbr} >> 1$, and $m_{cr} >> 1$.

\section*{Variance computation}

To simplify the computation of the variance of $\Delta$ under the interference
model in (\ref{eq:linear_interference}), we assume that there are sufficiently
many units in the completely randomized arm and sufficiently many clusters in
the cluster-based randomized arm such that the units and clusters can be
considered independently assigned to treatment and control, that the proportion
of treated units in each arm is $n_{cr,t}/n_{cr} = m_{cbr,t}/m_{cbr} = 1/2$, and
that $m_{cr} = m_{cbr}$, such that every cluster has an equal probability of
being assigned to either arm.
\begin{align*}
  \var_{\W, \Z} ( \Delta ) & = \E( \hat \tau_{cr}^2) + \E
  ( \hat \tau_{cbr}^2) - 2 \E (\hat \tau_{cr} \hat \tau_{cbr}) - \E(\hat \tau_{cr})^2 - \E(\hat \tau_{cbr})^2 + 2 \E(\hat \tau_{cr})
  \E( \hat \tau_{cbr}).
\end{align*}
Firstly, $\E_{\Z} \left( \hat \tau_{cr}^2 | \W\right)  = \sum_{i=1}^N W_i X_i +  \sum_{i=1}^N \sum_{j \neq i} W_i
  W_j X_{i,j}$, where
\begin{align*}
  X_i & = \E_\Z \left\{ \frac{1}{(n_{cr,t}^2)^{Z_i} (n_{cr,c}^2)^{1- Z_i}} \left(
  \beta^2 Z_i^2 + 2 \beta \frac{\gamma}{|\N_i|} Z_i \sum_{p \in \N_i} Z_p +
  \frac{\gamma^2}{|\N_i|^2} \sum_{\N_i^2} Z_p Z_q \right) \middle| \substack{\W
  \\W_i = 1 }\right\},\\
  X_{i,j} & = \E_\Z \left\{ \frac{(-1)^{1 - Z_i + 1 - Z_j}}{n_{cr,t}^{Z_i + Z_j}
  n_{cr,c}^{(1- Z_i) + ( 1 - Z_j)}} \left( \beta^2 Z_i Z_j + 2 \beta
  \frac{\gamma}{|\N_j|} Z_i \sum_{p \in \N_j} Z_p + \frac{\gamma^2}{|\N_i|
  |\N_j|} \sum_{\substack{p \in \N_j \\ q \in \N_i}} Z_p Z_q \right) \middle|
  \substack{ \W \\ W_i = 1 \\ W_j = 1} \right\}.
\end{align*}
Recall that $\calC(i)$ is the set of units that belong to the same cluster as
unit $i$ and that $\N_i$ is the neighborhood of unit $i$ in the graph. Consider $A_i = |\{p, q \in \N_i \cap C(i)\}||\N_i|^{-2}$, $B_i = |\{p, q \in
\N_i : C(p) \neq C(q)\}||\N_i|^{-2}$, and $C_i = |\{ p, q \in \N_i \backslash
C(i) : C(p) = C(q)\}||\N_i|^{-2}$, then
\begin{align*}
  \E_{\W}(W_i X_i) &= \frac{4}{N^2} \left\{ \beta^2 + 2\beta\gamma +
  \gamma^2 \left( A_i + B_i + \frac{3}{2} C_i \right) \right\}.
\end{align*}
Similarly, $\E( W_i W_j X_{ij}) = N^{-2} \left(2 \{i \in \calC(j)\}  + \{i
\notin \calC(j)\} \right) \left( \beta^2 + \frac{\gamma^2}{|\N_i||\N_j|}
\right)$, where $\{i \in \calC(j)\}$ is the boolean variable, equal to $1$ if
$i$ is in unit j's cluster. To compute $\E_{\W, \Z}( \hat
\tau_{cbr}^2)$, consider
\begin{align*}
  X_i & =\frac{16}{N^2} \E_\Z \left( \beta^2 Z_i^2 +
  \frac{2 \beta\gamma Z_i}{|\N_i|} \sum_{p \in \N_i} Z_p + \frac{\gamma^2}{|\N_i|^2}
  \sum_{\N_i^2} Z_p Z_q  \middle|\substack{ \W \\  W_i = 0 } \right). \\
  X_{i,j} & = \frac{16}{N^2} \E_\Z \left\{(-1)^{- Z_i - Z_j} \left( \beta^2
  Z_i Z_j + \frac{2 \beta \gamma Z_i}{|\N_j|} \sum_{p \in \N_j} Z_p +
  \frac{\gamma^2}{|\N_i| |\N_j|} \sum_{p \in \N_j, q \in \N_i} Z_p Z_q \right)
  \middle| \substack{ \W \\ W_i = 0 \\ W_j = 0} \right\}.
\end{align*}
Letting $\rho_i = |\N_i \cap \calC(i)||\N_i|^{-1}$, $\E_W(W_i X_i) = 4 N^{-2}
\left\{ \beta^2 + \beta \gamma \left( 1 +
  \rho_i \right) + \gamma^2 \left(A_i + B_i/2 + 3C_i/4 \right) \right\}$. For the
  cross-terms belonging to the same cluster $\calC(i) = \calC(j) = \calC(i,j)$,
  we introduce the following quantities:
\begin{align*}
  D_{ij} & = \frac{|\{p \in \N_j \cap \calC(i, j), q \in \N_i \backslash C(i,j)
  \} \cup \{p \in \N_j \backslash \calC(i, j), q \in \N_i \cap
  \calC(i,j)\}|}{|\N_i||\N_j|}, \\
  E_{ij} &= \frac{|\{p \in \N_j \cap \calC(i,j), q \in \N_i \cap
  \calC(i,j)\}|}{|\N_i||\N_j|}.
\end{align*}
It follows that $\E\{W_i W_j X_{ij} | i \in \calC(j)\} = 4 N^{-2}\left\{ \beta^2
+ \beta \gamma \left(1 + \rho_i\right) + \gamma^2 \left(D_{ij}/2 + E_{ij}
\right) \right\}$. For the cross-terms belonging to different clusters, we
introduce the following quantity:
\begin{equation*}
  F_{ij} =\frac{|[p \in \N_j \cap \{ C(j) \cup C(i)\}, q \in \N_i \cap \{C(i) \cup
    C(j)\} :  C(p) \neq C(q)]|}{|\N_i| |\N_j|}.
\end{equation*}
It follows that $\E_\W(W_i W_j X_{ij}) = \frac{1}{N^2} \left(\beta^2 +  \beta
\gamma \rho_i +\gamma^2F_{ij} \right)$. The cross-arms cross-terms are $\E_{\W, \Z} (\hat \tau_{cr} \hat \tau_{cbr}) =  \sum_{i, j} W_i (1 - W_j)
  X_{ij}$, where $X_{ij}$ is equal to
\begin{equation*}
 \frac{16}{N^2} \E_\Z \left\{(-1)^{- Z_i - Z_j} \left( \beta^2
  Z_i Z_j + \frac{\beta \gamma Z_i}{|\N_j|} \sum_{p \in \N_j} Z_p + \frac{\beta
  \gamma Z_j}{|\N_i|} \sum_{p \in \N_i} Z_p  + \frac{\gamma^2}{|\N_i| |\N_j|}
  \sum_{\substack{p \in \N_j\\ q \in \N_i}} Z_p Z_q \right) \middle|
  \substack{\W \\ W_i = 0 \\W_j = 1} \right\}.
\end{equation*}
It follows that $\E_{\W, \Z} (\hat \tau_{cr} \hat \tau_{cbr}) = N^{-2} \sum_{i,
j} \{\calC(i) \neq \calC(j)\}(\beta^2 + \beta \gamma \rho_i)$. Finally, the square
and the product of the expectations of our estimators are $\E_{\W, \Z}(\hat
\tau_{cr})^2 = \beta^2$, $\E_{\W, \Z}(\hat \tau_{cbr})^2 =\left(\beta + \gamma
\rho_C \right)^2$, and $\E_{\W, \Z}(\hat \tau_{cr}) \E_{\W, \Z}(\hat \tau_{cbr})
= \beta \left(\beta + \gamma \rho_C \right)$.  Let $\bar A = N^{-1} \sum_{i=1}^N A_i$, $\bar B = N^{-1} \sum_{i =1}^N
B_i$, $\bar C = N^{-1} \sum_{i=1}^N C_i$, $\bar D = N^{-2} \sum_{i \neq j} \{C(i)
= C(j)\} D_{ij}$, and  $\bar E = N^{-2} \sum_{i\neq j} \{ C(i) = C(j)\} E_{ij}$,
$\bar F = N^{-2} \sum_{i,j} \{C(i) \neq C(j)\} F_{ij}$, and finally $\bar G =
N^{-2} \sum_{i\neq j} |\N_i|^{-1}|\N_j|^{-1}$. Putting this all together, we
conclude that:
\begin{equation*}
  \var_{\W, \Z}(\Delta) \approx \beta^2 \left(\frac{8}{N} + \frac{5}{2M} \right)
  - \beta\gamma \rho_C  + \gamma^2 \left( \frac{8 \bar A}{N} + \frac{6 \bar
  B}{N} +\frac{9 \bar C}{N} +  \bar G +  \bar F - \rho_C^2  \right).
\end{equation*}
\section*{Proof of Theorem~\ref{thm:br_vs_cr}}
In order to compare the variance of the difference-in-means estimator under the
completely randomized assignment to its variance under a Bernoulli randomized
assignment, we consider a \emph{re-randomized} Bernoulli assignment strategy which rejects any
assignment where all units are in treatment or in control.  Let $\eta_t$
(resp.~$\eta_c$) be the \emph{realized} number of units assigned to treatment
(resp.~control) under the re-randomized Bernoulli assignment, and $n_t$ be the
desired number of units assigned to treatment under the completely randomized
assignment. Naturally, $\E(\eta_t) = n_t$.  Let $p \defeq n_t/N \in (0,1)$,
where $N$ is the total number of units.  By Eve's law,
\begin{equation}\label{eq:eveslaw}
  \var_\Z ( \hat \tau) = \var_{\eta_t} \left\{ \E_\Z \left( \hat \tau | \eta_t
  \right) \right\} + \E_{\eta_t} \left\{ \var_\Z \left( \hat \tau | \eta_t
  \right)\right\}.
\end{equation}
The expectation of $\hat \tau$ conditional on $\eta_t$ is $\eta_t^{-1} \sum_i \E_\Z( Z_i | \eta_t) Y_i(1) - \eta_c^{-1}
\sum_i \E_\Z( 1 - Z_i | \eta_t) Y_i(0)$.  Since $\E_\Z( Z_i | \eta_t) =
\eta_t/N$ and $\E_\Z( 1 - Z_i | \eta_t) = \eta_c/N$, it follows that $\E_\Z(\hat
\tau | \eta_t)$ is a constant equal to $N^{-1} \sum_i \{Y_i(1) - Y_i(0)\}$, and
thus $\var_{\eta_t} \left\{ \E_\Z \left(\hat \tau \middle| \eta_t \right)
\right\} = 0$.  The second term of (\ref{eq:eveslaw}) is $S_t
\E_{\eta_t}\left(\eta_t^{-1} \right) + S_c \E_{\eta_c} \left( \eta_c^{-1}
\right) - S_{tc} N^{-1}$, where $S_t = \sigma^2\{Y(1)\}$ and $S_c =
\sigma^2\{Y(0)\}$.  Lemma~\ref{lem:negative_moment} provides an upper-bound of
$\E \left(\eta_t^{-1} \right)$.
\begin{lemma}\label{lem:negative_moment}
  If $p^N + (1-p)^N \leq N^{-2} \leq 1/4$, then $\left|
  \E_{\eta_t} \left(\eta_t^{-1}\right) -  n_t^{-1} \right| \leq 5 n_t^{-2}$.
\end{lemma}
It follows that, for $N \geq 2$, $\left|\var_{\Z \sim BR}(\hat \tau) - \var_{\Z
\sim CR}(\hat \tau) \right| \leq 5 \left(S_t n_t^{-2} + S_c n_c^{-2} \right)$.
%
\section*{Proof of Lemma~\ref{lem:negative_moment}}
If $X$ is a binomial of parameters $B(n,p)$, let $p_{k} \defeq \proba(X = k)$
and let $\alpha \defeq p^n + (1-p)^n$. Then, for all $k \in [1,\dots,n-1]$, the
probability of the total number of treated units $n_t$ is $\proba(\eta_t = k) =
p_k/(1 - \alpha)$.
\begin{align*}
  \forall k \in [1,n-1],~\proba(N_t = k) & = \sum_{i=0}^{+\infty} \proba(i^{th}
  \text{~throw} = k | \text{first $i-1$ throws}=0~\text{or~}1) \\
  &= \sum_{i =0}^{+\infty} p_k \prod_{j =0}^{i-1} (p^n + (1 - p)^n) \\
  &= p_k \sum_{i =0}^{+\infty} (p^n + (1 -p)^n)^i \\
  &= \frac{p_k}{1 - p^n - (1-p)^n} \\
  &= \frac{p_k}{1 - \alpha}
\end{align*}

As expected, $\eta_t$ behaves \emph{almost} like a binomial distribution when $n
\rightarrow + \infty$. There is a known closed form formula for the first
negative moment of a binomial distribution from~\citet{chao1972negative}. For a
binomial $X$ of parameters $(n,p)$, $\E_{X} \left\{(1 + X)^{-1} \right\} =
\{p(n+1)\}^{-1} \left\{ 1 - (1 -
  p)^{n+1} \right\}$. Let $X \sim B(n, p)$,
\begin{equation*}
  \left|  \E_{\eta_t} \left(\frac{1}{\eta_t}\right) -  \frac{1}{np} \right| =
  \left|\E_{\eta_t} \left(\frac{1}{\eta_t}\right) - \frac{1}{1-\alpha} \E\left(
  \frac{1}{1 + X} \right) \right|  + \left| \frac{1}{1-\alpha}
  \E\left( \frac{1}{1 + X} \right) -  \frac{1}{n p} \right|.
\end{equation*}
The first term is
\begin{align*}
 \left|\E \left(\frac{1}{X}\right) - \frac{1}{1-\alpha} \E\left(
 \frac{1}{1 + X} \right) \right|
   &= \frac{1}{1 - \alpha} \left|  \sum_{k = 1}^{n-1} \frac{p_k}{k} -
    \sum_{k=0}^{n} \frac{p_k}{k+1} \right| \\
    &= \frac{1}{1-\alpha} \left\{ \sum_{k=1}^{n-1} \frac{p_k}{k (k+1)} +
  \frac{p^n}{n+1} + (1-p)^n \right\},
\end{align*}
where an $O(n^{-2})$-upper-bound of the summation term is $\sum_{k=1}^{n-1}
p_k/\{k (k+1)\} \leq 3(np)^{-2}$. The second term is upper-bounded by $p^n(1 -
\alpha)^{-1}n^{-1}$:

\begin{align*}
  \sum_{k=1}^{n-1} \frac{p_k}{k (k+1)} &= \sum_{k=1}^{n-1} \frac{1}{k (k+1)}
  \binom{n}{k} p^k (1-p)^{n-k} \\
  & \leq \sum_{k=1}^{n-1} \frac{3}{(k+1) (k+2)}
  \binom{n}{k} p^k (1-p)^{n-k} \\
  & = \frac{3}{(n+1) (n+2)} \sum_{k=1}^{n-1} \binom{n+2}{k+2} p^k (1-p)^{n-k}
  \\
  & = \frac{3}{p^2(n+1)(n+2)} \sum_{k=1}^{n-1} \binom{n+2}{k+2} p^{k+2}
  (1-p)^{n-k} \\
  & = \frac{3}{p^2(n+1)(n+2)} \sum_{k=3}^{n+1} \binom{n+2}{k} p^{k}
  (1-p)^{n+2-k} \\
  & \leq \frac{3}{p^2(n+1)(n+2)}  \leq \frac{3}{(np)^2}
\end{align*}

As a result, if $\alpha \leq n^{-2} \leq 1/4$,
\begin{equation*}
  \left|  \E \left(\frac{1}{X}\right) -  \frac{1}{n p} \right| \leq
  \frac{1}{1-\alpha} \left\{ \frac{3}{(np)^2} + \frac{p^n}{n+1} + (1-p)^n +
  \frac{p^n}{n} \right\} \leq \frac{5}{n^2 p^2}.
\end{equation*}


\end{document}